\documentclass[conference]{IEEEtran}

\usepackage[T1]{fontenc}
\usepackage{graphicx}
\usepackage{cite}
\usepackage{picinpar}
\usepackage{amsmath}
\usepackage{amssymb,bm}
\usepackage{float}
\usepackage{stfloats}
\usepackage{url}
\usepackage[latin1]{inputenc}
\usepackage{colortbl}
\usepackage{soul}
\usepackage{multirow}
\usepackage{pifont}
\usepackage{color}
\usepackage{alltt}
\usepackage{enumerate}
\usepackage{siunitx}
\usepackage{hyperref} 
\usepackage{breakurl}
\usepackage{epstopdf}
\usepackage{pbox}
\usepackage[linesnumbered,ruled]{algorithm2e}
\usepackage{comment}
\usepackage{amsthm}
\theoremstyle{plain}

\usepackage{subfig}
\usepackage{lipsum}

\begin{document}

\title{Clustering-based Joint Channel Estimation and Signal Detection for Grant-free NOMA}

\author{\IEEEauthorblockN{Ayoob Salari\IEEEauthorrefmark{1},
		Mahyar Shirvanimoghaddam\IEEEauthorrefmark{1}\IEEEauthorrefmark{2}, Muhammad Basit Shahab\IEEEauthorrefmark{2}, Reza Arablouei\IEEEauthorrefmark{3}, Sarah Johnson\IEEEauthorrefmark{2}}
\IEEEauthorblockA{\IEEEauthorrefmark{1}School of Electrical and Information Engineering, The University of Sydney, NSW 2006, Australia \\
\IEEEauthorrefmark{2}School of Electrical Engineering and Computing, The University of Newcastle, NSW 2308, Australia \\
\IEEEauthorrefmark{3}CSIRO's Data61, Pullenvale QLD 4069, Australia\\
Email: ayoob.salari@sydney.edu.au, mahyar.shm@sydney.edu.au, basit.shahab@newcastle.edu.au,\\ reza.arablouei@csiro.au, sarah.johnson@newcastle.edu.au
}}


\maketitle

\begin{abstract}
We propose a joint channel estimation and signal detection technique for the uplink non-orthogonal multiple access using an unsupervised clustering approach. We apply the Gaussian mixture model to cluster received signals and accordingly optimize the decision regions to enhance the symbol error rate (SER). We show that when the received powers of the users are sufficiently different, the proposed clustering-based approach with no channel state information (CSI) at the receiver achieves an SER performance similar to that of the conventional maximum likelihood detector with full CSI. Since the accuracy of the utilized clustering algorithm depends on the number of the data points available at the receiver, the proposed technique delivers a tradeoff between the accuracy and block length. 
\end{abstract}


\IEEEpeerreviewmaketitle

\section{Introduction}\label{introduction}
\IEEEPARstart{W}{ith} the dramatic growth of the Internet of things (IoT) applications and services, new communication strategies are warranted to provide fast, reliable, and scalable connectivity for a diverse range of IoT scenarios. IoT applications have various service requirements in terms of throughput, reliability, availability, end-to-end latency, energy/resource efficiency, security/privacy, and communication range~\cite{borza2019design,lavassani2018combining,shahab2019grantfree}. Cellular networks are considered to provide the network infrastructure for a large proportion of future IoT applications due to their wide coverage, scalability, and security features~\cite{shirvanimoghaddam2017massive}. Accordingly, the release 15 and above of the 3rd generation partnership project (3GPP) have devised particular solutions for massive IoT, ultra-reliable low-latency communications (URLLC), and industrial IoT (IIoT)~\cite{5GNREricsson}. 

One of the main challenges for massive IoT applications is that the number of devices is usually large while the available spectrum is limited. To tackle this, non-orthogonal multiple access (NOMA) has emerged as a promising technology that allows multiple users to simultaneously transmit their data over the same radio resource~\cite{shahab2019grantfree}. Some prominent NOMA techniques are power-domain NOMA, sparse code multiple access (SCMA), multi-user shared access (MUSA), and interleave division multiple access (IDMA)~\cite{vaezi2019multiple}. In power domain NOMA, users share the same channel and are distinguished at the receiver based on their different power levels. In SCMA, users' data are mapped to multi-dimensional codewords, which not only have the ability to suppress the inter-user interference, but also achieve a diversity gain~\cite{shahab2019grantfree}. In MUSA, complex-valued sequences are used for spreading the users' data enabling the system to handle more users. In IDMA, repetition coding and interleavers are applied to the users' data streams to reduce the inter-user interference when combining them~\cite{lin2018optimal}.

In NOMA, multiuser detection techniques such as joint user detection or successive interference cancellation (SIC) are usually applied at the receiver to decode messages. Assuming perfect channel state information (CSI) at the base station (BS), the multi-user bit error rate (BER) performance for power-domain uplink NOMA is investigated in~\cite{wang2017closed,yeom2019ber,kara2018BER}.
Joint user detection techniques typically outperform SIC in terms of BER~\cite{yeom2019ber}. However, they incur much higher complexity at the receiver side. Attaining perfect CSI at the BS requires a significant amount of time within the channel coherence interval \cite{mirza2019optimization}, which can considerably limit the throughput and increase the overhead for massive IoT applications with short packet communications~\cite{Shirvani2019Short}.

Main channel estimation methods for obtaining CSI can be categorized as training-based, blind, and semi-blind~\cite{zhang2006optimal,shin2007blind,shaham2019fast,liu2013semi,ladaycia2019semi}. At the expense of a lower throughput, the training-based estimation methods can accurately attain CSI with a relatively low complexity when long training sequences are available \cite{zhang2006optimal}. On the other hand, blind estimation methods make use of the properties of the transmitted signal to estimate the channel without any training symbol. Although the blind estimation methods are more bandwidth efficient, they are generally less accurate compared with the training-based ones~\cite{shin2007blind}. Semi-blind estimation methods utilize the merits of both training-based and blind methods to create a balance between throughput and accuracy~\cite{liu2013semi}. Channel estimation is particularly challenging in NOMA as the BS needs to estimate the CSI of multiple users and the receiver error performance is severely degraded due to channel estimation errors~\cite{rim2019uplink}. 

Grant-free NOMA has been proposed to reduce the signaling overhead and enhance the access capability so that all potential users can freely access the channel without waiting for any scheduling grant~\cite{shahab2019grantfree}. Grant-free NOMA significantly reduces the overhead of control signaling and helps meet the requirements of massive IoT. However, it poses challenges for reliable receiver design. A straightforward approach to the receiver design is to first identify the active users, then estimate their channel coefficients, and finally recover their transmitted data. However, this separate processing approach may consume substantial time and power resources, which in turn degrades the system performance~\cite{jiang2019joint}.


Recently, the problem of joint estimation and detection in grant-free NOMA has received increasing attention~\cite{jiang2019joint,wang2016joint,wei2016approximate}. Authors of~\cite{chen2018sparse,liu2018massive,zhang2018block} reformulate the joint channel estimation and user detection problem such that the pilot symbols suit a compressed sensing problem. Due to the underlying sparsity, the number of training symbols is reduced~\cite{jiang2020joint}. However, in massive IoT, since the packets are small, even a few training symbols can lead to a major efficiency loss. 

In this paper, we propose a method for joint channel estimation and user detection in uplink NOMA without using any training symbol. We employ an unsupervised machine-learning algorithm to cluster the received signals at the receiver side. Using the clustering results, we estimate the channel and perform SIC to detect each user. We show that when the powers of the signals received from the users are sufficiently different, the proposed clustering-based method with no CSI at the receiver achieves the same performance in terms of symbol error rate as the conventional maximum-likelihood detector with full CSI. We also show that the proposed method offers a tradeoff between accuracy and block length as the performance of the clustering depends on the number of data points (symbols) available at the receiver.

The rest of the paper is organized as follows. Section II presents the system model and provides some preliminary information. The proposed clustering method based on Gaussian mixture models is detailed in Section III. Numerical results are provided in Section IV. Finally, Section V concludes the paper. 

%
%
%
%
%
%
%
%
%
%
%
%
%

\section{System Model and Preliminaries}  
\label{System Model}
We consider an uplink massive IoT system where devices use NOMA to share the available radio resources. In particular, we consider $K$ frame-synchronized single-antenna users transmitting their data to the BS using the quadrature phase shift keying (QPSK) modulation. 
%
%
Let us denote the channel between the $u$th user and the BS by $h_u$ and assume it to be a zero-mean circular symmetric complex Gaussian random variable, i.e., $h_u\sim\mathcal{CN}(0,\beta_u)$ where $\beta_u$ represents the large-scale fading component including path-loss and shadowing. We assume block fading, that is $h_u$ remains unchanged for each transmission frame of length $N$ symbols. 

The received superimposed signal vector at the BS, denoted by $\mathbf{y}\in \mathbb{C}^{N\times 1}$, can be expressed as 
\begin{equation} \label{eq:1}
    \mathbf{y} = \mathbf{X}\mathbf{h}  + \mathbf{n}
\end{equation} 
where $\mathbf{X}=[\mathbf{x}_1,\cdots,\mathbf{x}_K]$ represents the matrix of transmitted symbols,  $\mathbf{x}_u\in \mathbb{C}^{N\times 1}$ is the length-$N$ message transmitted from user $u\in \{1,\cdots,K\}$, $\mathbf{h}\in \mathbb{C}^{K\times 1}$ contains the channels $h_u$ for all users,  $\mathbf{n}\sim\mathcal{CN}(0, \nu\mathbf{I}_N)$ is the multivariate additive white Gaussian noise, $\mathbf{I}_N$ is the $N\times N$ identity matrix, and $\nu$ is the noise power. We further assume that $h_u$, $\mathbf{x}_u$, and $\mathbf{n}$ are statistically independently of each other. 

The user symbols $x_{u,i}$ are modulated using a common modulation scheme with the signal constellation $\mathcal{S}$ and cardinality $|\mathcal{S}|$, i.e., each $x_{u,i}$ is randomly and uniformly drawn from $\mathcal{S}$. For simplicity, we assume that all users use QPSK to transmit their messages, i.e., $x_{u,i} \in \{\exp({\frac{-j\pi}{4}}), \exp({\frac{-j3\pi}{4}}), \exp({\frac{j\pi}{4}}), \exp({\frac{j3\pi}{4}})\}$ $\forall u\in\{1,\cdots,K\}$ and $i\in\{1,\cdots,N\}$\footnote{The proposed clustering-based joint channel estimation and signal detection technique can be easily extended to consider more complex constellations.}. The signal-to-noise ratio (SNR) for the $u$th user is then given by $\gamma_u=\beta_u/\nu$.
It is easy to show that the entries of $\mathbf{y}$, denoted by $y_i$, are i.i.d. and have the following Gaussian mixture distribution
\begin{align}
    y_i\sim\frac{1}{|\mathcal{S}|^K}\sum_{\mathbf{s}^{(i)}\in{\mathcal{S}}^{K}}\mathcal{CN}\left(\mathbf{h}^T\mathbf{s}^{(i)},~\nu\right).
    \label{eq:GMMmodel}
\end{align}


Throughout the paper, we assume that all users are frame synchronized, which can be achieved by frequently sending beacon signals from the BS. We also assume that the BS does not know the CSI to any user. Therefore, it attempts to jointly estimate the channels and detect the signals. However, we assume that the BS knows the number of transmitting users, $K$, and the modulation type of each user.


\section{The Proposed clustering-based Joint Channel Estimation and Signal Detection Technique}  
\label{ML} 
\begin{figure*}[t] 
\subfloat[Single user, $\gamma = 12$dB \label{F4a}]{%
\includegraphics[width=0.66\columnwidth]{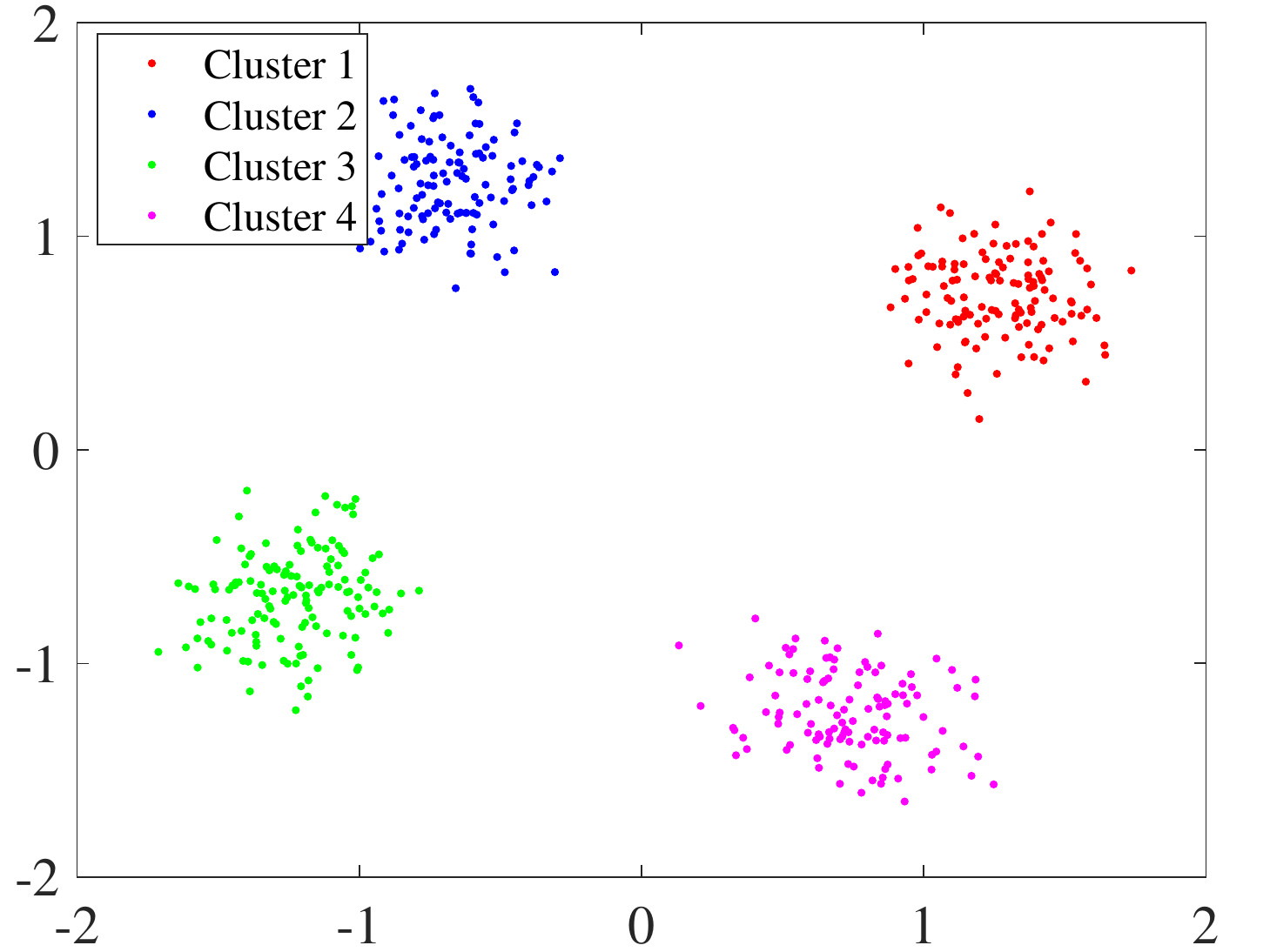}
}
\hfill
    \subfloat[2-user NOMA, $\gamma_1 =20$dB, $\gamma_2 =17$dB   \label{F4b}]{%
\includegraphics[width=0.66\columnwidth]{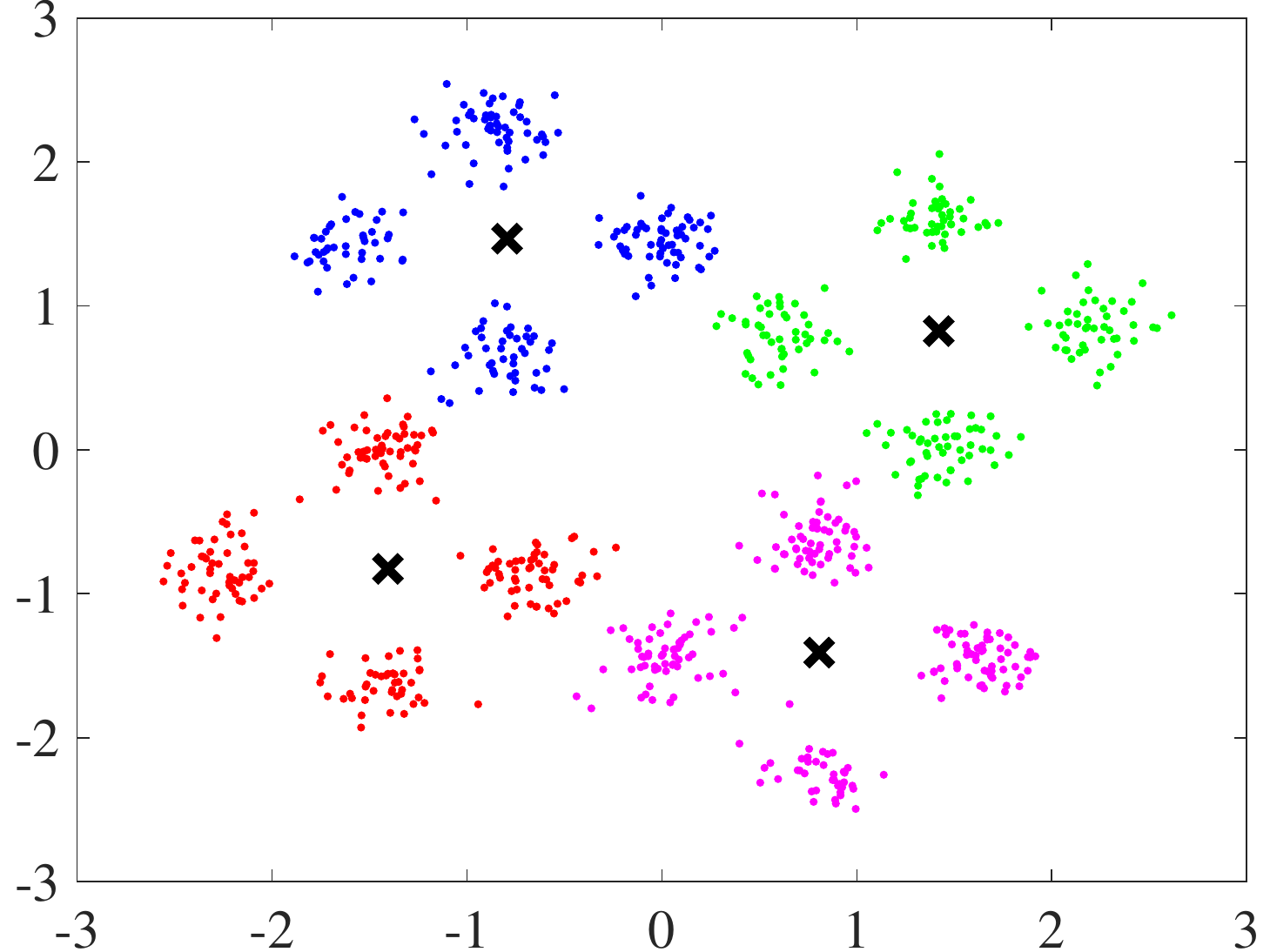}
}
\hfill
\subfloat[2-user NOMA, $\gamma_1 =10$dB, $\gamma_2 =7$dB  \label{F4c}]{%
\includegraphics[width=0.66\columnwidth]{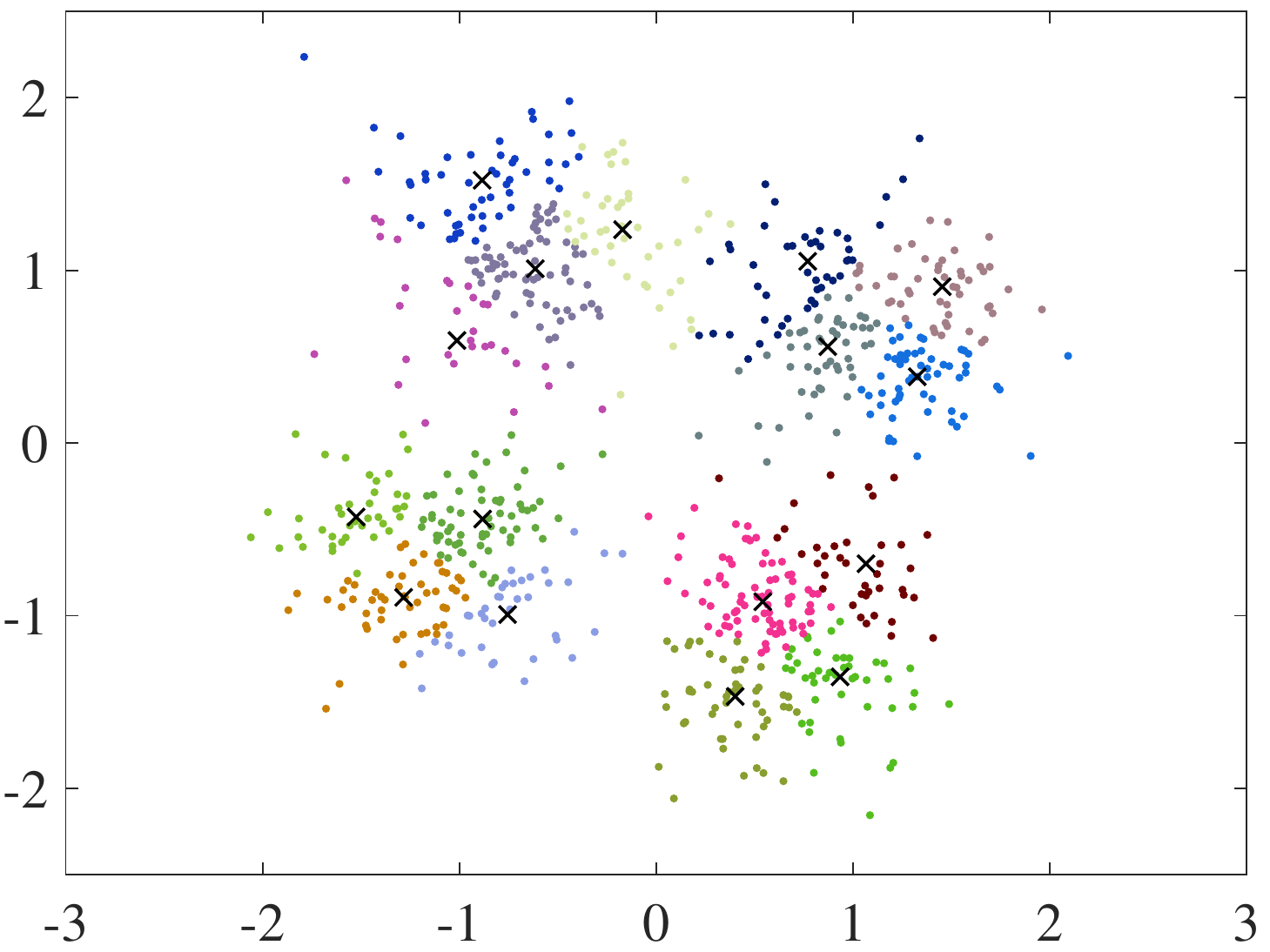}
}
\caption{\small{Received signal constellation diagram at BS, when $N=500$.}}
\label{fig:constellation}
\vspace{-1.5em}
\end{figure*}

We first provide a few examples to better illustrate the basic concepts of the proposed approach. Fig. \ref{fig:constellation} shows the signal points in the I-Q plane collected at the receiver for a point-to-point communication system (Fig. \ref{fig:constellation}a) and two-user NOMA  (Fig. \ref{fig:constellation}b,c) when both users utilize QPSK modulation to transmit their messages. As seen in Fig.\ref{F4a}, with a single active user, signal points are clumped together into four clusters, whose centroids can be used to estimate the amplitude and phase of the channel. Similarly, in Fig. \ref{F4b}, the data points of two users can be clustered into 16 clusters as the BS knows the number of active users. For this particular example where the clusters are separated from each other, the channel between each user and the BS can be estimated accurately. However, when the clusters overlap due to high noise and fading, as in Fig. \ref{F4c}, estimating the channels and detecting the individual signals are more challenging. In what follows, we propose an effective approach to cluster the received signals and perform joint channel estimation and signal detection. 


%

There are several clustering algorithms, such as, K-means, DBSCAN \cite{ester1996density}, OPTICS \cite{ankerst1999optics,kriegel2011density}, mean shift \cite{cheng1995mean}, and Gaussian mixture model (GMM) \cite{singh2009statistical}, that can be used to estimate the clusters of the received signals. Since all users use the same modulation scheme and the channels are block fading, the clusters formed at the BS are symmetric with roughly the same densities (see Fig. \ref{fig:constellation}). Moreover, as the noise has Gaussian distribution, the received signal can be modeled by a mixture of Gaussian distributions as in (\ref{eq:GMMmodel}), which makes GMM a natural choice for our clustering problem. In our experiments, we have found the GMM clustering to be more effective compared with the above-mentioned alternatives, especially when the users have comparable received powers at the BS or when the SNR is low.
\subsection{GMM-based Clustering for the SIC receiver}
In SIC, the receiver first decodes the signal of the strongest user (the user with the highest power), subtract it from the combined received signal, then decodes the next strongest signal and so on.
To apply SIC at the receiver of our system model, we initially divide the received signals into four clusters representing User 1's signals (data points). Then, we divide each of those clusters into four extra clusters representing user 2's signals and so on. Taking the received signals as our observed data, the considered joint channel estimation and signal detection problem boils down to estimating the unknown latent parameters of the assumed Gaussian mixture distribution in \eqref{eq:GMMmodel}.

Since we deal with complex signals, we denote a  $2$-dimensional multivariate Gaussian probability density function by $g(\mathbf{z};\bm{\mu},\bm{\Sigma})$ where $\bm{\mu}$ and $\mathbf{\Sigma}$ are the mean vector and the covariance matrix, respectively, and express it as
\begin{equation}\label{eq:3}
      g(\mathbf{z};\bm{\mu},\bm{\Sigma}) = \frac{\exp{ \left(-\frac{1}{2} (\mathbf{z} -\bm{ \mu})^T \mathbf{\Sigma}^{-1} (\mathbf{z} - \bm{\mu})  \right)}}{\sqrt{(2\pi)^{2} |\mathbf{\Sigma}|}}.  
\end{equation}
%

In GMM clustering, the number of clusters is known and the data is assumed to be generated by a mixture of Gaussian distributions. A GMM parameterizes the mean, covariance, and weight of each Gaussian distribution component. When a common $M$-ary modulation scheme is adopted by all users, there are $M$  Gaussian distributions each with weight $\omega_j$, $j\in\{1,\cdots,M\}$. Accordingly, the underlying Gaussian mixture distribution can be written as a weighted sum of the $M$ constituting Gaussian distributions (each representing a cluster), i.e.,
\begin{equation}\label{eq:4}
       p(\mathbf{z};\bm{\mu}_1,...,\bm{\mu}_M, \bm{\Sigma}_1,...\bm{\Sigma}_{M}) = \sum_{j=1}^{M} \omega_j g_j(\mathbf{z};\bm{\mu}_j,\mathbf{\Sigma}_j)
 \end{equation}
where $\sum_{j=1}^M\omega_j=1$. 
%
We are interested in estimating $\bm{\mu}_j$, $\mathbf{\Sigma}_j$, and $\omega_j$, $j=1,\cdots, M$, from the observed data. This can be done by maximizing the likelihood function  (\ref{eq:4}) for all received signals. 
%
%
To this end, we utilize the expectation maximization (EM) algorithm \cite{hastie2009elements}, which is suitable for solving maximum likelihood problems with unobserved latent variables.

Let $\Delta_{i,j}$ symbolize the association of the $i$th data point to the $j$th cluster represented by the $j$th Gaussian distribution. Therefore, we have
$$ \Delta_{i,j}= \begin{cases} 1 \hspace{5mm} \text{if} \hspace{1mm} \mathbf{z}_i \hspace{1mm} \text{belongs to the cluster} \hspace{1mm} g_j\\
0 \hspace{5mm} \text{otherwise}. \end{cases} $$
It is clear that $P(\Delta_{i,j} = 1)  = \omega_j$ and 
$P(\Delta_{i,j} = 0)  = 1 - \omega_j$. However, both $\Delta_{i,j}$ and $\omega_j$ are unknown. 
%
In the $t$th iteration of the EM algorithm, we first estimate the so-called responsibility variable of each model $j$ for every observation $i$ defined as
\begin{align}\label{eq:7}
        \hat{\gamma}_{i,j}^{(t)} = \frac{\hat{\omega_j}^{(t-1)} g_j(\mathbf{z}_i;\hat{\bm{\mu}}_j^{(t-1)}, \hat{\mathbf{\Sigma}}_j^{(t-1)})}{\sum_{k=1}^{{M}} \hat{\omega_{k}}^{(t-1)} g_{k}(\mathbf{z}_i;\hat{\bm{\mu}}_{k}^{(t-1)}, \hat{\mathbf{\Sigma}}_{k}^{(t-1)})}.
 \end{align}
We then assign each data point to its corresponding cluster. In particular, for each $\mathbf{z}_i$ we find $m^{(t)}_i=\arg\max_j\hat{\gamma}^{(t)}_{i,j}$ and set 
$$ \Delta^{(t)}_{i,j}= \begin{cases} 1 \hspace{5mm} \text{if} \hspace{1mm} j=m^{(t)}_i\\
0 \hspace{5mm} \text{otherwise}. \end{cases} $$

We define the corresponding log-likelihood function as
\begin{align}\label{eq:5}
       &l^{(t)} (\bm{\mu}_1,\cdots,\bm{\mu}_{M}, \bm{\Sigma}_1,\cdots,\bm{\Sigma}_{M}|\mathbf{z}_1,\cdots,\mathbf{z}_N) \\ \nonumber
       & = \sum_{i=1}^{N} \left[ \sum_{j=1}^{M} \Delta_{i,j}^{(t)} \ln \left( \omega_j^{(t)} g_j(\mathbf{z}_i;\bm{\mu}_j^{(t)},\mathbf{\Sigma}_j^{(t)}) \right)  \right].
 \end{align}
%
%
In the next step of the EM algorithm, we use the calculated responsibilities to update the mean, variance, and weight of each cluster as
\begin{align}\label{eq:8}
       \hat{\omega}_j^{(t)}  &= \frac{ \sum_{i=1}^{N} {\hat{\gamma}}_{i,j}^{(t)}}{ \sum_{i=1}^{N} \sum_{k=1}^{{M}}  {\hat{\gamma}}_{i,k}^{(t)} }, \\
       \label{eq:9}
       \hat{\bm{\mu}}_j^{(t)}  &= \frac{ \sum_{i=1}^{N} \hat{\gamma}_{i,j}^{(t)} \mathbf{z}_i }{ \sum_{i=1}^{N}  \hat{\gamma}_{i,j}^{(t)} },\\
       \label{eq:10}
       \hat{\mathbf{\Sigma}}_j^{(t)}  &= \frac{ \sum_{i=1}^{N} \hat{\gamma}_{i,j}^{(t)} (\mathbf{z}_i - \hat{\bm{\mu}}_j^{(t)}) (\mathbf{z}_i - \hat{\bm{\mu}}_j^{(t)})^T }{ \sum_{i=1}^{N}  \hat{\gamma}_{i,j}^{(t)} }. 
 \end{align}

After convergence, $m_i$, $i=1,\cdots,N$, contain the final clustering results. The EM algorithm is guaranteed to converge to a local optimum \cite{dempster1977maximum}.  

When the signals of the users are uniformly drawn from the same QPSK constellation, the weights of the Gaussian distributions are the same, i.e., $\omega_j = \frac{1}{4}$, $j=1,\cdots,4$. Moreover, using the QPSK modulation and a SIC receiver, at each stage of the SIC, we need to estimate only four Gaussian distributions. This helps with managing the computational complexity. 

The proposed approach is summarized in Algorithm \ref{GMM Clustering Algorithm}. This algorithm is applied at each iteration of SIC. In other words, for a two-user scenario, we first run the algorithm to detect the four clusters of user 1, then we run it again to identify user 2's clusters.
\begin{figure*}[t] 
\subfloat[$N=500$ \label{fig:P2P-500symbol}]{%
\includegraphics[width=0.66\columnwidth]{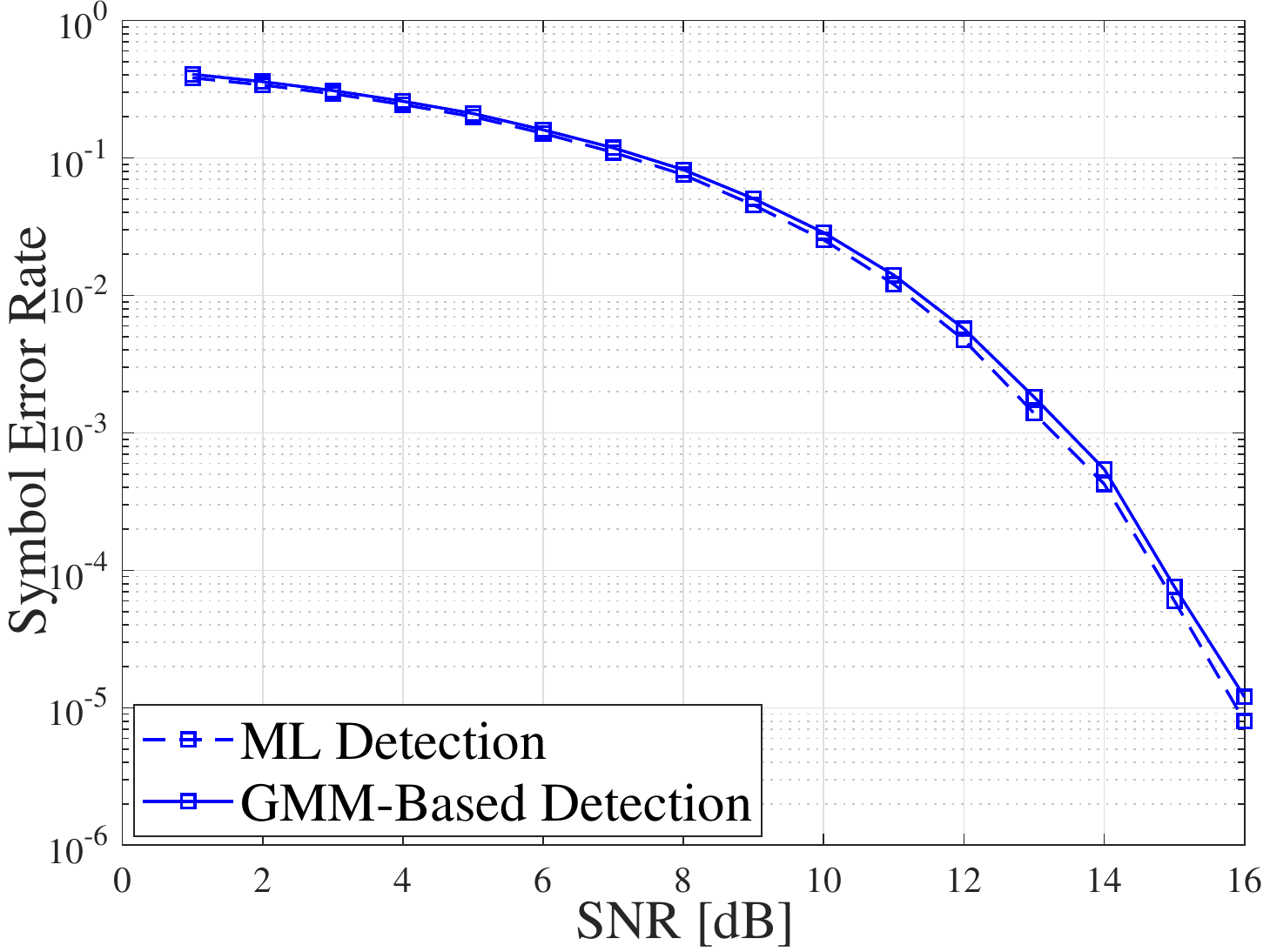}
}
\hfill
    \subfloat[$N=100$  \label{fig:P2P-100symbol}]{%
\includegraphics[width=0.66\columnwidth]{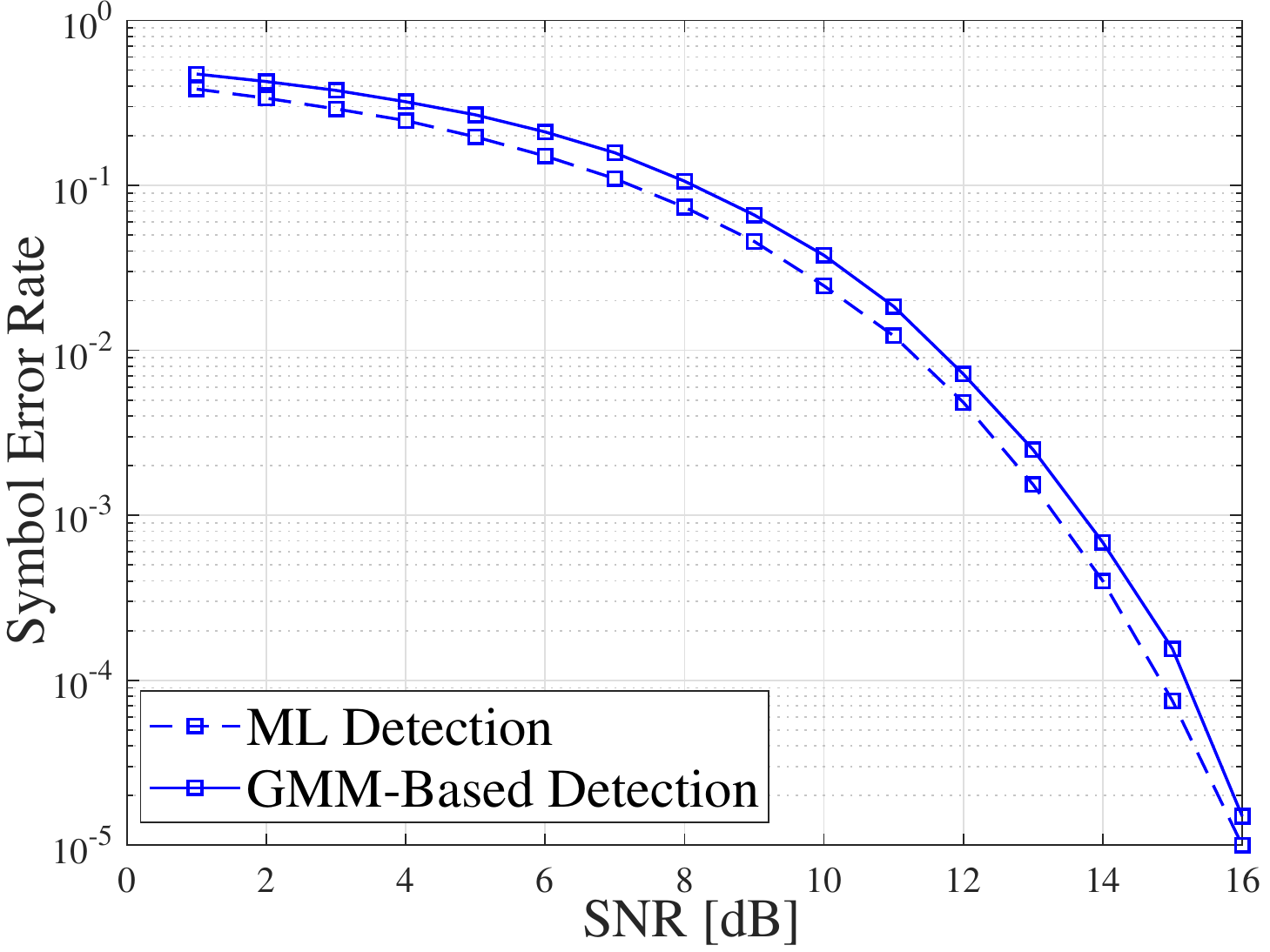}
}
\hfill
\subfloat[$N=50$ \label{fig:P2P-50symbol}]{%
\includegraphics[width=0.66\columnwidth]{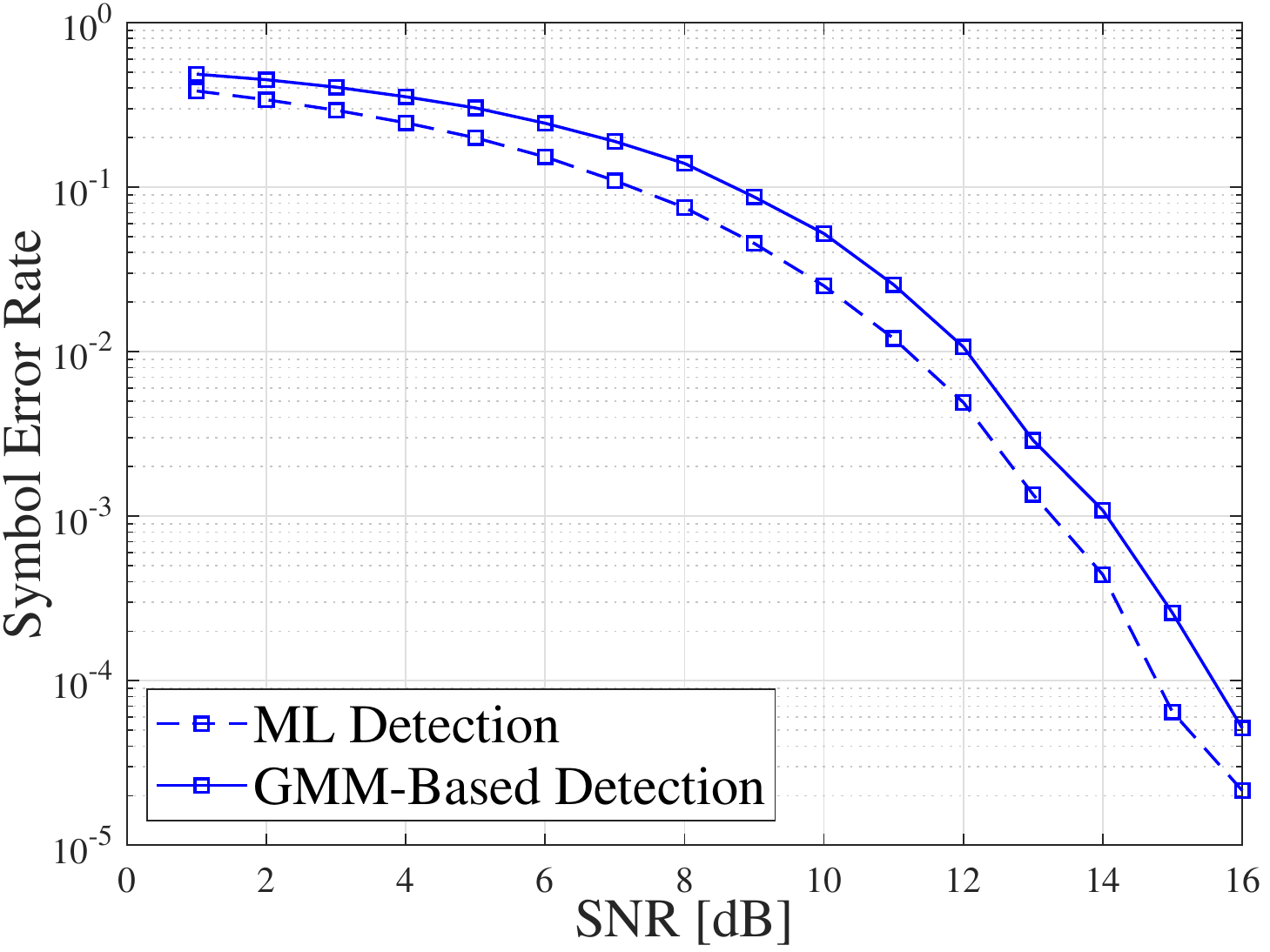}
}
\caption{\small{SER Comparison of GMM Clustering and optimal ML Detection Optimal for point-to-point communication.}}
\label{fig:P2P simulation}
\vspace{-1.5em}
\end{figure*}

In the initialization step, we divide the received data into four quadrants and calculate the mean and covariance of each cluster. Then, we fix the weights and calculate the responsibility and log-likelihood function, according to (\ref{eq:7}) and (\ref{eq:5}), respectively. We then apply the EM algorithm to find four cluster centers and covariance matrices (lines 6 to 11 in Algorithm \ref{GMM Clustering Algorithm}). Next, we calculate the phase of each cluster center. Since we consider QPSK modulation, each cluster center has a $\frac{\pi}{2}$ phase difference from the adjacent clusters. Due to the phase rotation caused by the noise and channel fading, the center phase is not exactly $\frac{\pi}{4}, \frac{3\pi}{4}, \frac{5\pi}{4}$ or $\frac{7\pi}{4}$. To minimize the effect of phase rotation, we calculate the phase difference between each center and their expected value. Afterwards, we average the phase rotations as $\theta = [({\phi_1} -\frac{\pi}{4})  + ({\phi_2} -\frac{3\pi}{4}) + ({\phi_3} -\frac{5\pi}{4}) + ({\phi_4} -\frac{7\pi}{4})]/4$ (Step 13). Given the average phase rotation, the decision boundaries are updated accordingly. The channel gain can also be found by taking an average over the vectors representing the centroids. 

\begin{algorithm}[t]
        \KwIn{Received data at BS, number of Gaussian distributions M, and convergence threshold $\epsilon$}
        \KwOut{Clustered data, mean (centroid) of each cluster $\hat{\mathbf{\mu}}$ and covariance of each cluster  $\hat{\mathbf{\Sigma}}$.}
            \textbf{Initialize} $\hat{\mathbf{\mu}}^{(0)}$ and $\hat{\mathbf{\Sigma}}^{(0)}$ by dividing received data into four quadrants 
            
            Set $\omega_j = 0.25$
            
            Calculate $\hat{\gamma}_{i,j}^{(0)}$ according to (\ref{eq:7})
            
            Calculate log-likelihood function according to (\ref{eq:5})
            
            Set $t=1$
           
        \While {$ l^{(t)} - l^{(t-1)} \geq \epsilon $}
            {
            Update $\hat{\bm{\mu}}^{(t)}$ and $\hat{\mathbf{\Sigma}}^{(t)}$ using (\ref{eq:9}) and (\ref{eq:10})
            
            Update $\hat{\gamma}_{i,j}^{(t)}$ according to (\ref{eq:7})
            
            Update log-likelihood function according to (\ref{eq:5})
            }
        \textbf{Return} optimal $\hat{\mathbf{\mu}}$ and $\hat{\mathbf{\Sigma}}$
        
        Calculate the phase of each cluster centroid ($\phi_i$)
        
        Calculate the mean phase rotation as  $\theta = \frac{\sum_{i=1}^{M} {\phi_i} - 4\pi}{4} $
        
        Update the QPSK decision boundaries based on the phase rotation
        \caption{Applying GMM for Clustering}
        \label{GMM Clustering Algorithm}
\end{algorithm}

\section{Numerical Results and Discussion}
\begin{figure*}[t] 
\subfloat[$N=500$ \label{fig:NOMA-2USer-500symbol-3db}]{%
\includegraphics[width=0.66\columnwidth]{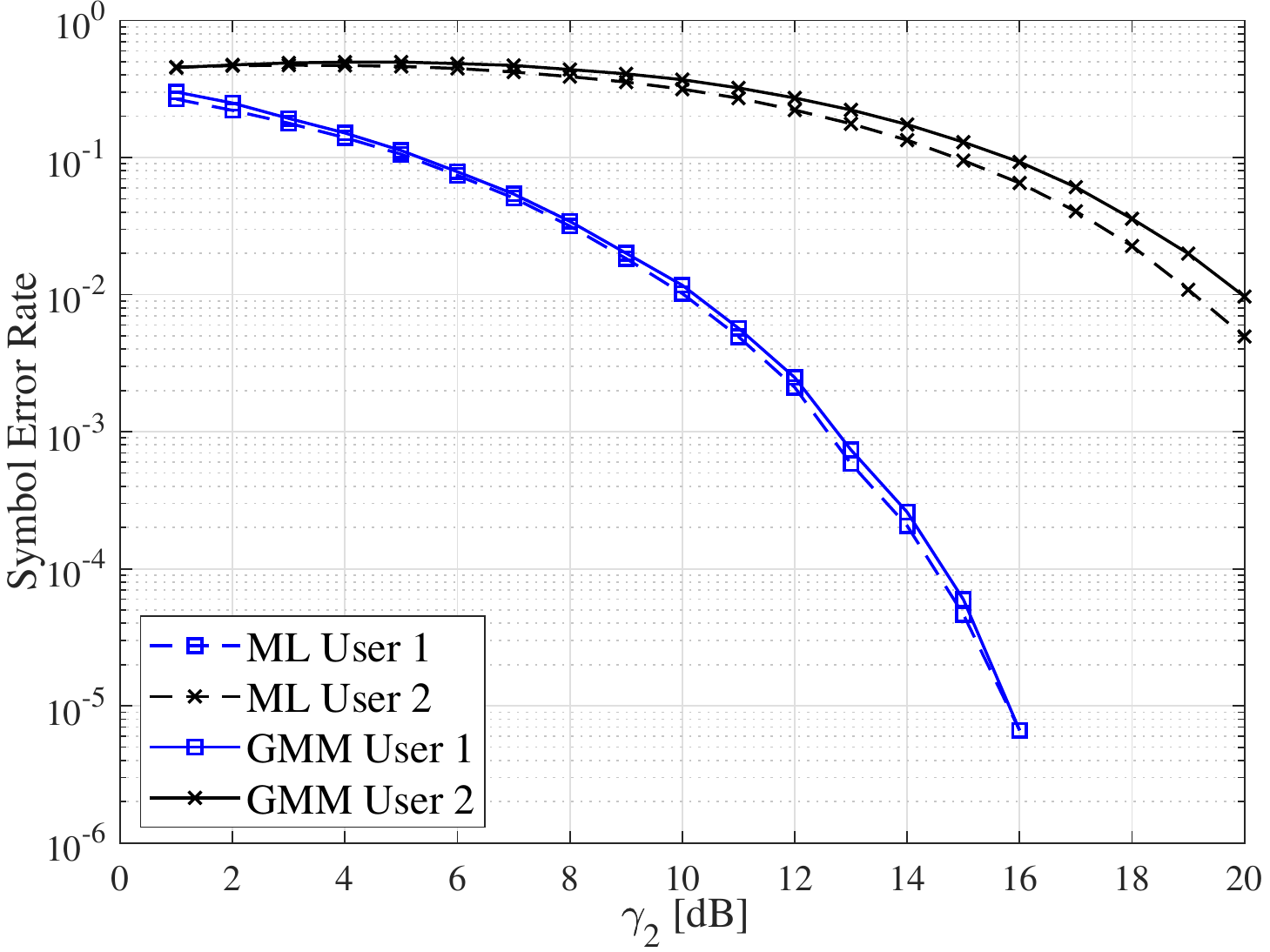}
}
\hfill
    \subfloat[$N=100$  \label{fig:NOMA-2USer-100symbol-3db}]{%
\includegraphics[width=0.66\columnwidth]{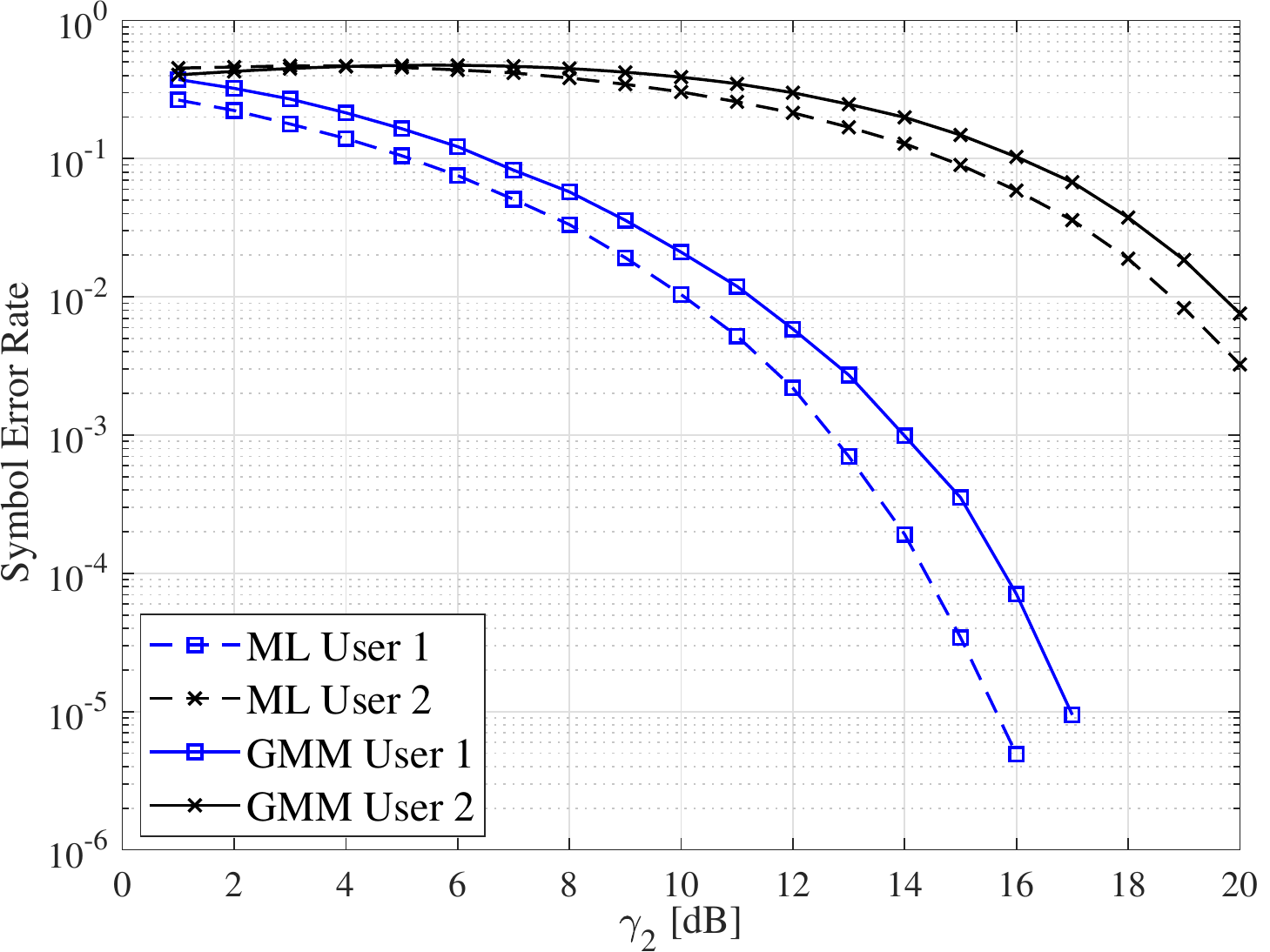}
}
\hfill
\subfloat[$N=50$ \label{fig:NOMA-2USer-50symbol-3db}]{%
\includegraphics[width=0.66\columnwidth]{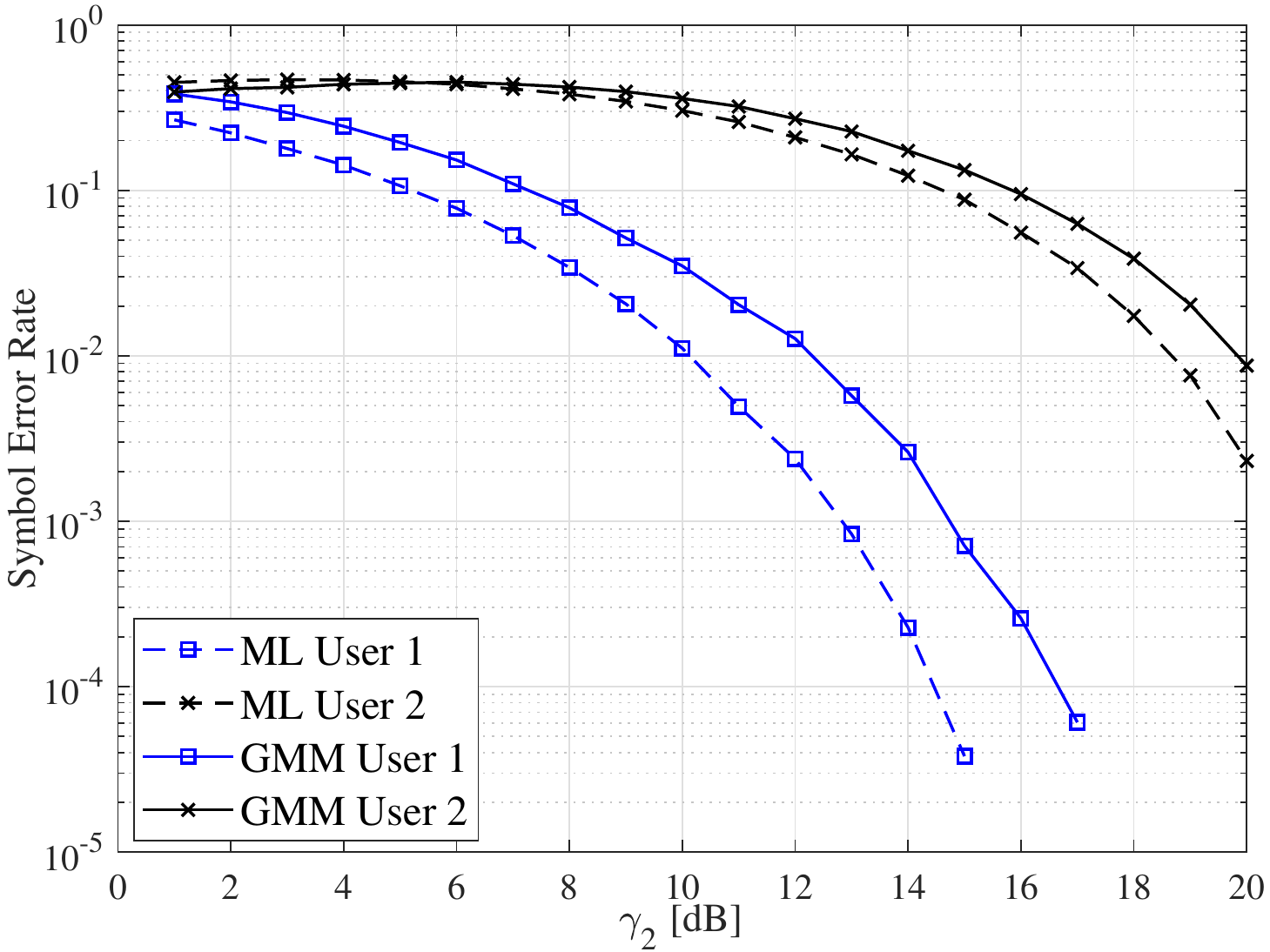}
}
\caption{\small{SER comparison of the GMM-clustering-based and ML techniques for two-user NOMA when $\gamma_1-\gamma_2=3$dB.}}
\vspace{-1.5em}
\label{fig:2 users NOMA 3db}
\end{figure*}
\begin{figure*}[t] 
\subfloat[$N=500$ \label{fig:NOMA-2USer-500symbol-6db}]{%
\includegraphics[width=0.66\columnwidth]{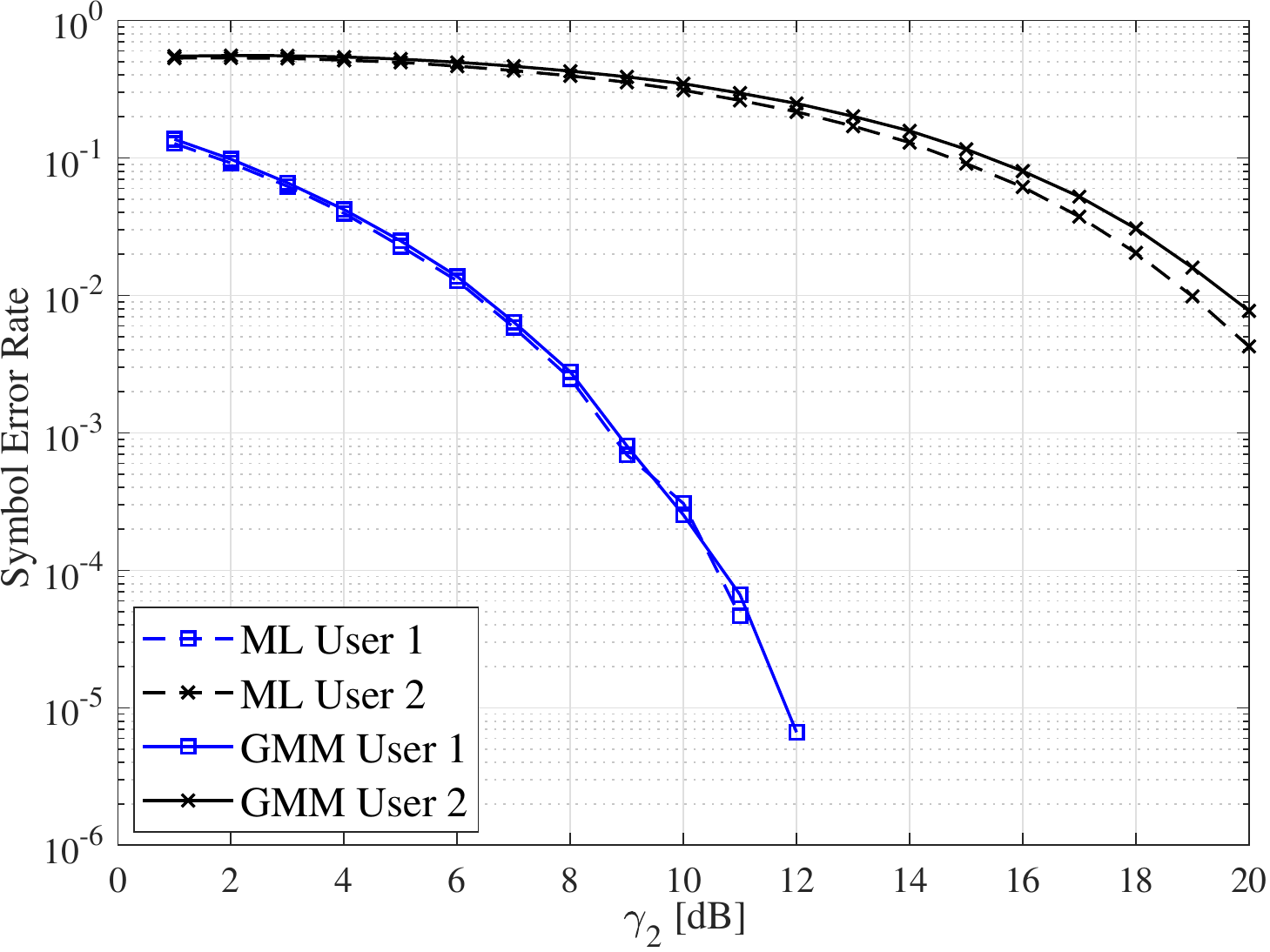}
}
\hfill
    \subfloat[$N=100$  \label{fig:NOMA-2USer-100symbol-6db}]{%
\includegraphics[width=0.66\columnwidth]{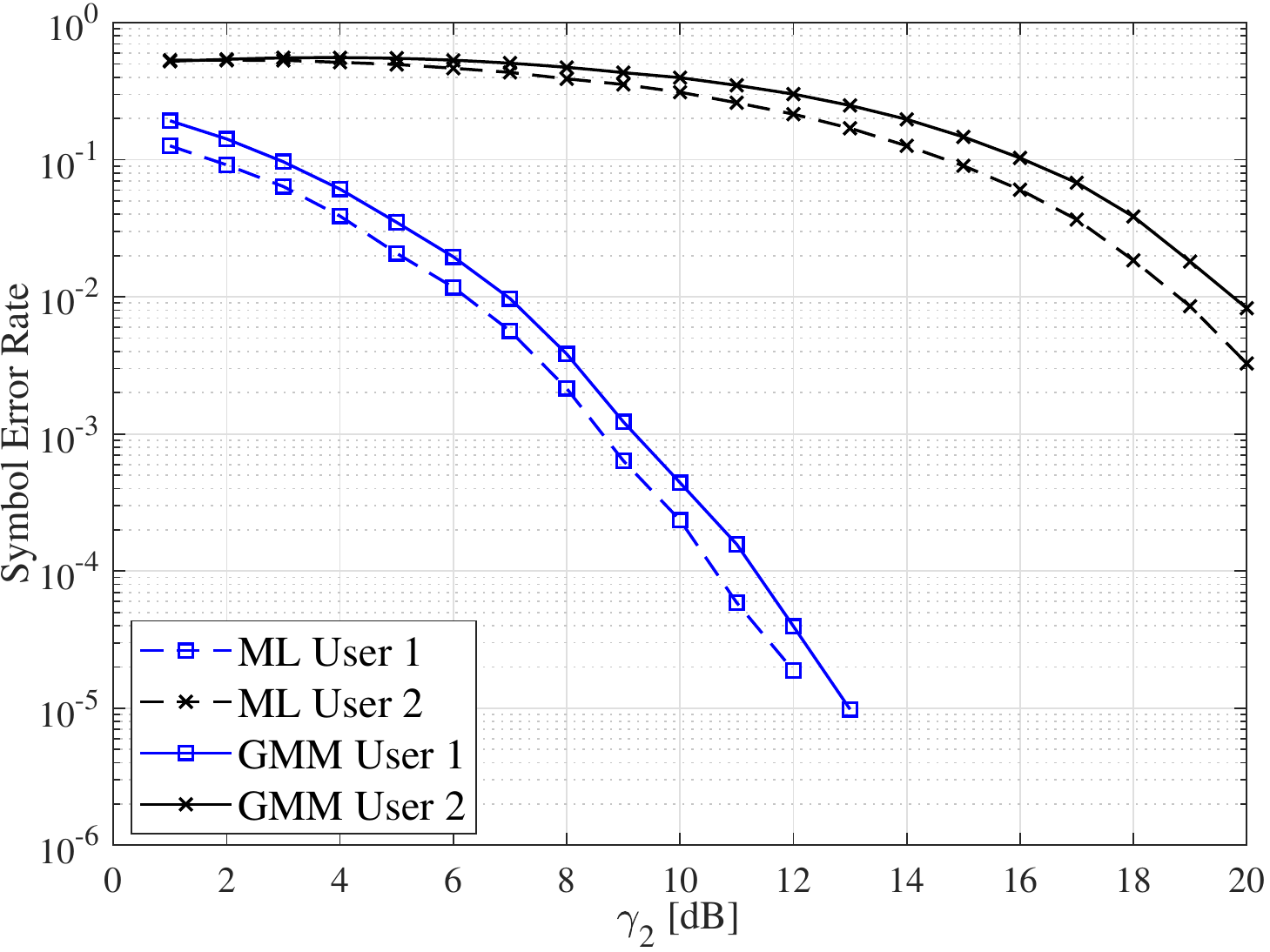}
}
\hfill
\subfloat[$N=50$ \label{fig:NOMA-2USer-50symbol-6db}]{%
\includegraphics[width=0.66\columnwidth]{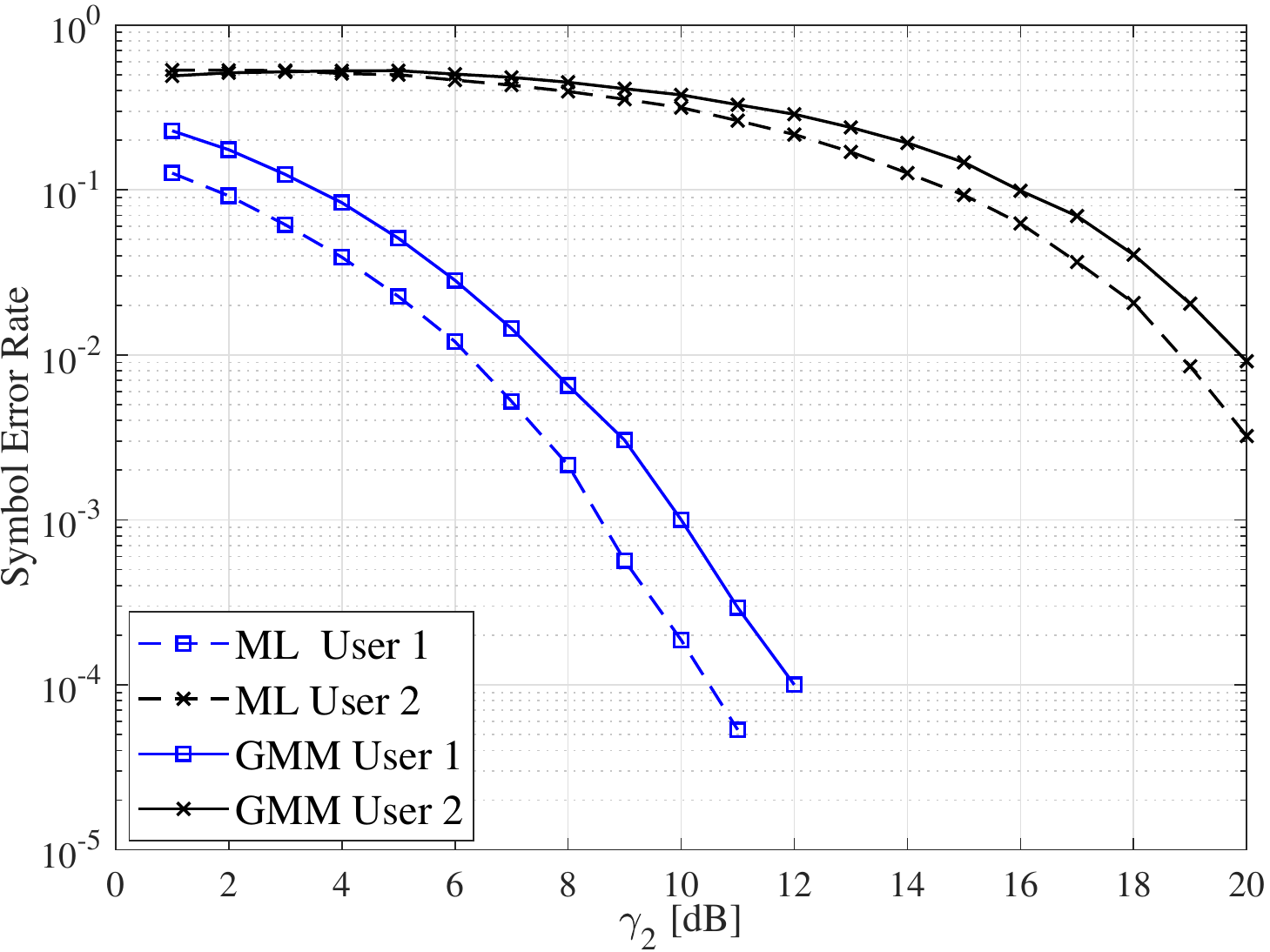}
}
\caption{\small{SER comparison of the GMM-clustering-based and ML techniques for two-user NOMA when $\gamma_1-\gamma_2=6$dB.}}
\label{fig:2 users NOMA 6db}
\vspace{-1.5em}
\end{figure*}
In this section, we compare the symbol error rate (SER) performance of the proposed GMM-based clustering approach for the SIC receiver with the SER performance of the maximum-likelihood (ML) receiver where the BS has full CSI hence is able to attain the optimal decision boundaries. 
%
%

Fig. \ref{fig:P2P simulation} shows the SER versus SNR for a point-to-point communication scenario when the users use QPSK modulation. It is clear from this figure that when the sample size $N$ is sufficiently large, the proposed technique performs very close to the optimal ML detection with full CSI. Although the accuracy increases, the complexity also increases. When the number of samples is small (Fig. \ref{fig:P2P simulation}c), there is a rather small difference in the performance of the two methods that can mainly be attributed to the sub-optimal decision boundaries found by GMM due to limited observations.

Fig. \ref{fig:2 users NOMA 3db} shows the performance of the proposed GMM clustering approach for a two-user NOMA scenario when the users have 3dB power difference, i.e. $\gamma_1-\gamma_2=3$dB. As it can be seen in this figure, the proposed GMM-clustering-based technique can accurately determine the clusters and performs symbol detection with an SER very close to that of the optimal ML detection with full CSI. Similar to the point-to-point scenario, when the sample size $N$ is small, the performance deviates from the optimal. However, the proposed technique is still advantageous since, unlike the ML detection technique, it does not require any pilot symbol or separate training phase to estimate the CSI. This is however achieved with an increased complexity at the receiver.  Communicating any training symbol sequence is generally inefficient when the number of symbols (the packet size) is small. One needs to send at least six symbols to acquire (semi-)accurate CSI to each user at the receiver \cite{Shirvani2019Short}. This results in 12\% loss in throughput when the packet contains 50 symbols, which leads to further reduction in throughput for a multi-user scenario.

\begin{figure}[t]
    \centering
    \includegraphics[width=\columnwidth]{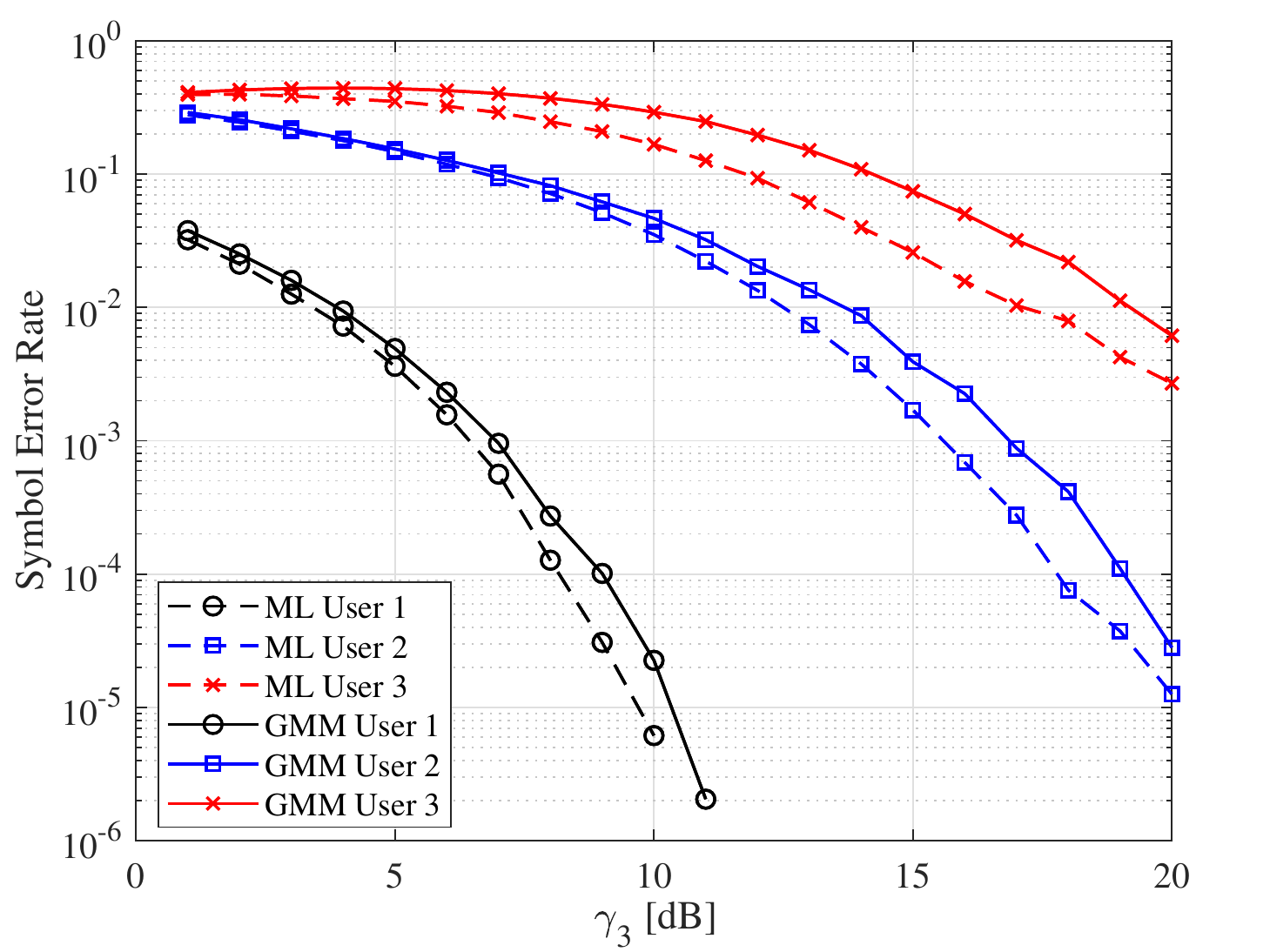}
    \caption{\small{SER comparison of the GMM-clustering-based and ML techniques for three-user NOMA when $\gamma_1-\gamma_2=6$dB,  $\gamma_2-\gamma_3=6$dB, and $N=500$.}}
    \label{fig:3userNOMA}
    \vspace{-1.5em}
\end{figure}

To further investigate the effectiveness of the proposed technique, in Fig. \ref{fig:2 users NOMA 6db}, we provide the results for the two-user NOMA scenario when the users have a $6$dB power difference, i.e., $\gamma_1-\gamma_2=6$dB. As it can be observed in this figure, the SER gap between the proposed and optimal ML detection techniques is smaller when users have a larger power difference. Fig. \ref{fig:3userNOMA} shows the SER performance of a three-user NOMA when the number of symbols is 500. It is seen that the proposed technique can accurately detect the signals.

It is important to note that Step 13 of Algorithm 1 can be further enhanced to improve the phase detection. In particular, when performing symbol detection for User 2, the proposed technique will obtain 16 phase values representing the clusters around 4 initial clusters of User 1. In this case, one may use all 16 estimated phases to obtain a more accurate estimation of the User 2's channel. The proposed technique can also be modified to exploit a few training symbols to improve the performance. That is instead of sending training sequences for each user to estimate the channels individually for ML detection, the users can simultaneously send a few training symbols to help improve the cluster formation in the proposed technique. 


Since we assume that the noise affecting the received signals is i.i.d. additive white Gaussian, from a theoretical point of view, one may conclude that the GMM clustering used in our technique is equivalent to the well-known $K$-means clustering algorithm. However, in practice, the noise covariance matrix is not strictly a multiple of the identity matrix, i.e., the noise effecting different received signals may be correlated or have different variances. Therefore, the GMM clustering is more accurate than the $K$-means clustering as, unlike $K$-means, it does not assume the same covariance for all clusters but estimates the relevant covariance matrices for each cluster.

\section{Conclusion}
In this paper, we proposed a Gaussian mixture model (GMM) clustering-based joint channel estimation and signal detection technique for grant-free NOMA. In particular, we applied GMM in each iteration of the successive interference cancellation scheme to cluster constellation points. We then performed joint channel estimation and signal detection. The proposed approach does not rely on any training sequence to perform channel estimation, which makes it favorable over the maximum likelihood detection that requires full channel state information at the receiver. Simulation results showed that when the number of transmitted symbols is moderate or large the symbol error rate performance of the proposed technique is on a par with that of the optimal maximum likelihood detection. Since the accuracy of the employed clustering algorithm depends on the sample size, we demonstrated the existence of a tradeoff between the accuracy and the block length.

\bibliographystyle{IEEEtran}
\bibliography{IEEEabrv,main}

\end{document}